\documentclass[useAMS,usegraphicx]{mn2e}

\title[Prediction GeV gamma emission from young radio lobes]
{New prediction of extragalactic GeV $\gamma$-ray emission 
from radio lobes of young AGN jets}
\author[M. Kino, H. Ito, N. Kawakatu, H. Nagai]
{M. Kino,$^{1,2}$ 
H. Ito,$^{3}$ 
N. Kawakatu$^{1}$ and
H. Nagai$^{1}$\\ 
$^{1}$ National Astronomical Observatory of Japan, 181-8588 Mitaka, Japan\\
$^{2}$ ISAS/JAXA, 3-1-1 Yoshinodai, Sagamihara, Kanagawa 229-8510, Japan\\
$^{3}$ Department of Science $\&$ Engineering, Waseda University, 
Tokyo 169-8555, Japan}
\begin{document}

\date{}

\pagerange{\pageref{firstpage}--\pageref{lastpage}} \pubyear{2008}

\maketitle

\label{firstpage}

\begin{abstract}

We present a new prediction of GeV $\gamma$-ray 
emission from radio lobes of young AGN jets.
In the previous work of Kino et al. (2007), 
MeV $\gamma$-ray bremsstrahlung emission was predicted from
young cocoons/radio-lobes 
in the regime of no coolings.
In this study, we include cooling effects of bremsstrahlung emission
and adiabatic loss.
With the initial conditions determined by observed young
radio lobes, we solve a set of equations describing 
the expanding lobe evolution.
Then we find that 
the lobes initially have electron temperature of $\sim$GeV,
and they cool down to $\sim$MeV by the adiabatic loss.
Correspondingly, the lobes initially yield  bright bremsstrahlung
luminosity in $\sim$GeV range and they fade out.
We estimate these $\gamma$-ray emissions and
show that  nearby young radio lobes
could be detected with
Fermi Gamma-ray Space Telescope.

\end{abstract}

\section{Introduction}

Jets in active galactic nuclei (AGNs) are
one of the most powerful objects in the universe.
AGNs with jets are known as radio-luminous sources
and they have been called as radio-loud AGNs.
There is increasing evidence that radio-loud AGNs significantly
interact with ambient medium and
the cosmological importance of radio-loud AGN
feedbacks have been advocated (e.g., Best et al. 2007).
Among a variety of radio-loud AGNs,
a population with its linear size $LS<1~{\rm kpc}$
has been called as compact symmetric objects (CSOs).
The previous studies
support a scenario in which CSOs propagate 
from $\sim 100~{\rm pc}$  scales 
thrusting away an ambient medium and they grow up to 
Fanaroff-Riley type II radio galaxies (FR IIs)
(e.g., Fanti et al. 1995; 
Readhead et al. 1996a, b;
O'Dea \& Baum 1997; 
Stanghellini et al. 1998; 
Snellen et al. 2000;
Dallacasa et al. 2000; 
Orienti et al. 2007). 
CSOs are thus recognized as newly born AGN jets,
and they are crucial sources 
to explore physics of radio-loud AGNs in their early days.

On the contrary to various radio observations,
high-energy emission observations
of CSOs had been sporadic (Siemiginowska 2008 for review).
Recently, XMM-Newton observations 
begin to detect
X-ray emissions from some of the CSOs 
although these are not very bright in X-ray
(Guainazzi et al. 2006; Vink et al. 2006; O'Dea et al. 2006).
Although a few authors recently indicate possible $\gamma$-ray emissions
from CSOs 
(e.g., Stawarz et al. 2008),
there is little observations of CSOs viewed in $\gamma$-rays.
Hence $\gamma$-ray observations of CSOs 
may provide us new independent knowledges on the youth era
of AGN jets.

Recently, Kino et al. (2007) (hereafter KKI07) showed 
a possibility of
thermal MeV $\gamma$-ray bremsstrahlung emission 
 from the young radio-loud AGNs. 
In KKI07, as a first step, they focused on 
the simplest case in which cooling effects are not significant.
These sources roughly correspond to
medium-size symmetric objects (MSOs) with
$1~{\rm kpc}<LS<10~{\rm kpc}$ and FR II galaxies
with $LS > 10~{\rm kpc}$.
Since various cooling timescales in smaller sources tend to be 
shorter than the ones in larger sources,
cooling effects in CSOs are expected to be 
more significant than the ones in MSOs and FR IIs.
However little is known about the cooling processes
in compact radio sources such as CSOs. 
In this study we will consistently solve a set of equations
which describe young lobe expansions
including the effects of bremsstrahlung emission
and adiabatic loss 
together with initial conditions determined by observations.

The goals of this paper are 
(i) to solve a hydrodynamical evolution of expanding lobes 
with the cooling effects
and 
(ii) to show a  new  prediction of $\gamma$-ray emission from the lobes
based on the results  of (i). 
In \S 2, we show a set of basic equations which describes an
expanding radio lobe inflated by its internal energy. 
In \S 3, we show the resultant temperature evolution of radio lobes.
New predictions of bright 
GeV $\gamma$-ray emissions from young radio lobes 
is presented in \S 4.
Summary and discussion are given in \S 5.

\section{Model of radio lobes}

\subsection{Basic equations}

Let us consider 
the dynamics of sideways expansion for young
radio lobes.
We assume that
the sideways expansion of radio lobe is caused by 
the internal pressure of the shocked jet matter.
A breakdown of the internal pressure is an open question.
The partial pressure of non-thermal electrons alone
might be small to expand the radio lobes
(e.g., Hardcastle and Worral 2000;
Ito et al. 2008 and reference therein).
Therefore the partial pressure of thermal particles
may play an important role for lobe expansions.
Here we focus on the case of thermally expanding lobes
and we neglect the partial pressure of non-thermal electrons for simplicity.

The equation of motion and 
the energy equation, 
for expanding radio lobes,
are respectively
given by
\begin{eqnarray}
\frac{d}{dt}\left(\frac{4\pi}{3}\rho_{\rm ext}R^{3}{\dot R}\right)
=4\pi R^{2}P  ,
\end{eqnarray}
\begin{eqnarray}
\frac{d}{dt}\left(\frac{4\pi}{3}R^{3}U_{\rm th}\right)
=L_{\rm j}-P\frac{d}{dt}\left(\frac{4\pi}{3}R^{3}\right)
-\frac{4\pi}{3}R^{3}\epsilon_{\rm cool}   ,
\end{eqnarray}
where 
$\rho_{\rm ext}$
$R$,
${\dot R}=v$,
$P$,
$U_{\rm th}(=3P)$, and
 $L_{\rm j}$ 
are
the mass density of external ambient medium, 
the radius,
the expansion velocity,
thermal pressure,
the thermal energy density of relativistic particles
of the lobe, and 
the total kinetic power of the jet, respectively.
We denote 
the volume emissivity as $\epsilon_{\rm cool}$ and here we 
focus on the case of  $\epsilon_{\rm cool}=\epsilon_{\rm brem}$
where $\epsilon_{\rm brem}$ is the bremsstrahlung one.
In the case of relativistic jets,  $L_{\rm j}$ is given by
\begin{eqnarray}
L_{\rm j}=\Gamma_{\rm j}{\dot M}_{\rm j}c^{2}  ,
\end{eqnarray}
where
$\Gamma_{\rm j}$, and 
${\dot M}_{\rm j}$ are
the bulk Lorentz factor of the relativistic jet, and 
mass outflow rate of the jet, respectively.
The $L_{\rm j}$ is the ultimate source of
the lobe expansion.
The $L_{\rm j}$ is a free parameter in this model
and it is assumed to be constant in time.
Particular noteworthy compared with KKI07
is that we take cooling effects into account.
The equation of state (EOS) and mass conservation 
in the lobe are, respectively, 
\begin{eqnarray}
P=(n_{-}+n_{+})kT_{\pm}+n_{p} kT_{p}  ,
\end{eqnarray}
\begin{eqnarray}
\frac{4\pi}{3}\rho R^{3}= {\dot M}_{\rm j}t  ,
\end{eqnarray}
where 
$n_{-}$,
$n_{+}$, and 
$n_{p}$
are
number densities of thermal
electrons, thermal positrons and total protons in the lobe, 
respectively,
and
$\rho=(n_{-}+n_{+})m_{e}+n_{p}m_{p}$, 
$T_{\pm}$, and 
$T_{p}$ are
total mass density and temperatures of 
$e^{\pm}$ pairs and protons, respectively.
Merely for simplicity, we neglect  protons
i.e.,  $n_{p}\approx 0$ throughout this work. 
Although ``which plasma component is dynamically
dominated in AGN jets?'' is still a matter of debate,
previous works at least indicate the existence of copious
amount of $e^{\pm}$ pairs in jets
(Wardle et al. 1998; 
Reynolds et al. 1996;
Sikora and Madejski 2000; 
Kino and Takahara 2004;
Kataoka et al. 2008).
Combining the above equations all together,
we obtain the ordinary differential equation of
\begin{eqnarray}\label{eq:rk}
\frac{d^{3}R}{dt^{3}}
+\frac{11{\dot R}{\ddot R}}{R}
+\frac{12{\dot R}^{3}}{R^{2}}=
\frac{3}{4\pi \rho_{\rm ext}R^{4}} 
\left(L_{\rm j}-\frac{4\pi R^{3}\epsilon_{\rm brem}}{3}\right).
\end{eqnarray}
Setting $\epsilon_{\rm brem}$ as
for relativistic $e^{\pm}$ pairs 
(Eq. (22) in Svensson 1982),
then we obtain
\begin{eqnarray}
\frac{4\pi R^{3}\epsilon_{\rm brem} }{3}
&=&0.75\times 10^{11}~
\rho_{\rm ext}{\dot M}t(R{\ddot R}+3{\dot R}^{2}) \nonumber \\
&\times& \left[\ln 
\left(\frac{4.4 \rho_{\rm ext}R^{3}(R{\ddot R}+3{\dot R}^{2})}
{9{\dot M}t c^{2}}\right) 
+\frac{5}{4} \right]~{\rm erg~s^{-1}} . \nonumber \\
\end{eqnarray}
%
Given the initial values of  
$R$, 
${\dot R}$,
${\ddot R}$ at an initial time $t_{0}$,
we can solve the evolution of expanding lobes.
The initial temperature can be obtained via 
$P=\rho_{\rm ext}(R{\ddot R}+3{\dot R}^{2})/3$ and EOS.

The free paramter $L_{\rm j}$  can be normalized as
\begin{eqnarray}\label{eq:lkin}
L_{\rm j}=5.7 \times10^{44}
\left(\frac{\Gamma_{\rm j}}{10}\right)
\left(
\frac{{\dot M}_{\rm j}}{10^{-3}~{\rm M_{\odot}~yr^{-1}}}
\right)~{\rm erg~s^{-1}}
\end{eqnarray}
which is typical value for AGN jets.
In this work, we will examine the range of 
$5.7 \times 10^{43}~{\rm erg~s^{-1}}<L_{\rm j}<
5.7 \times 10^{46}~{\rm erg~s^{-1}}$,
the corresponding ${\dot M}_{\rm j}$ is 
$1\times 10^{-4}~{\rm M_{\odot}~yr^{-1}}< 
{\dot M}_{\rm j} <
1\times 10^{-1}~{\rm M_{\odot}~yr^{-1}}$.
The mass outflow rate ${\dot M}_{\rm j}$ is
the essential parameter to
determine $L_{\rm j}$.

\subsection{Initial conditions of lobes and jets}

Based on recent observations
of CSOs 
(e.g., O'Dea 1998; Stanghellini et al 1998; Dallacasa 2000;
Snellen et al. 2004), 
we set the initial conditions as follows.
Since physical quantities diverge at $R = 0$,
certain initial conditions of the lobe
at $t=t_{0}\neq 0$ are required for solving a set of equations.
In this work, we treat the evolutions of physical quantities 
for $t\ge t_{0}$ and we do not consider $t < t_{0}$.
The followings are two
initial conditions (with subscript of $0$) examined in this work;
\begin{eqnarray}
R_{0}=100~{\rm pc}, \quad v_{0}=0.1c, \quad
t_{0}=3.0 \times 10^{3}~{\rm yr} \quad({\rm Low}~P_{0}) , \nonumber\\
R_{0}=300~{\rm pc}, \quad v_{0}=0.2c, \quad
t_{0}=4.5 \times 10^{3}~{\rm yr} \quad({\rm High}~P_{0}) ,
\end{eqnarray}
and we set $dv/dt|_{t=t_{0}}=0$. 
This size and velocity is typical ones for 
CSOs 
(e.g., Stanghellini et al. 1998; 
Dallacasa et al. 2000; 
Polatidis and Conway 2003; Gugliucci et al. 2005, 2007;
Taylor et al. 2000; 
Nagai et al. 2006;
Kawakatu et al. 2008).
Note that  $R_{0}\approx v_{0}t_{0}$ holds here.
Since we here deal with small radio lobes, 
the external ambient medium is not intracluster medium (ICM) 
but interstellar medium (ISM).
By using them, we further set the initial conditions as 
\begin{eqnarray}\label{eq:initial2}
P_{0}=m_{p}n_{\rm ext}v_{0}^{2} ,
\quad 
kT_{\pm,0}=
\frac{m_{e}P_{0}V_{0}}{{\dot M}_{\rm j}t_{0}}, \quad
n_{-,0}=
\frac{{\dot M}_{\rm j}t_{0}}{2m_{e}V_{0}}, 
\end{eqnarray}
where $V_{0}=4\pi R_{0}^{3}/3$.
Hereafter we set the number density of external ambient medium
$n_{\rm ext}=1~{\rm cm^{-3}}$
which is typical for ISM (e.g., Readhead et al. 1996a and references within).

In principle, the upper limit of  $n_{-}$ can be constrained by 
the analysis of Faraday depolarization (Dreher et al. 1987).
However, the strong Faraday
depolarization observed in CSOs (Cotton et al. 2003)
are likely to be caused by
dense foreground matter such as narrow line region.
Therefore  $n_{-}$ in radio lobes of CSOs 
has not been clearly constrained.


\section{Temperature of radio lobes}

Fig. \ref{fig:kT} shows the evolution of 
$kT_{\pm}$ of expanding lobes which is obtained 
by solving Eq. (\ref{eq:rk}) with the initial condition of the ``Low-$P_{0}$'' case.
We examine various outflow rate of 
${\dot M}_{\rm j}=10^{-4}~{\rm M_{\odot}~yr^{-1}}$,
${\dot M}_{\rm j}=10^{-3}~{\rm M_{\odot}~yr^{-1}}$, 
${\dot M}_{\rm j}=10^{-2}~{\rm M_{\odot}~yr^{-1}}$, and
${\dot M}_{\rm j}=10^{-1}~{\rm M_{\odot}~yr^{-1}}$.
Corresponding kinetic power of injected jet into
the lobes are, respectively, 
$L_{\rm j}=5.7\times 10^{42}~{\rm erg~s^{-1}} $,
$L_{\rm j}=5.7\times 10^{43}~{\rm erg~s^{-1}} $,
$L_{\rm j}=5.7\times 10^{44}~{\rm erg~s^{-1}} $, and 
$L_{\rm j}=5.7\times 10^{45}~{\rm erg~s^{-1}}$.
The initiall temperature of the lobe $kT_{\pm,0}$ in the case of 
the Low $P_{0}$ case can be analytically evaluated as 
\begin{eqnarray}\label{eq:kT0}
kT_{\pm,0}
&=& 0.18~{\rm GeV}
\left(\frac{{\dot M}_{\rm j}}{10^{-3}~{\rm M_{\odot}yr^{-1}}}\right)^{-1}
\left(\frac{P_{0}V_{0}/t_{0}}{2.0\times 10^{46} {\rm erg~s^{-1}}}\right) , \nonumber \\
\end{eqnarray}
The resultant $kT_{\pm}$ evolutions 
obtained by Eq. (\ref{eq:rk}) can be well 
approximated by the simple relation of
$kT_{\pm}
\approx
kT_{\pm,0}
(t/t_{0})^{-1}
+\Gamma_{\rm j}m_{e}c^{2}/3$
for  $kT_{\pm,0}\ge \Gamma_{\rm j}m_{e}c^{2}/3$.
In the early-phase,  the first term $(t/t_{0})^{-1}$
is dominant for the case examined here.
Then $kT_{\pm}$ decreases proportional to $t^{-1}$.
The agreement of this approximation 
is due to the negligible bremsstrahlung cooling 
and dominant adiabatic loss in the early-phase.
We will show the reason in the next section.
In the late phase,
it is found that the temperature
asymptotically
approaches to 
$kT_{\pm}=
\Gamma_{\rm j}m_{e}c^{2}/3
=1(\Gamma_{\rm j}/10)~{\rm MeV}$. 
This clearly coincides with the prediction of ``MeV cocoon" by KKI07.
This implies that the final temperature of 
lobes is determined only by $\Gamma_{\rm j}$
regardless of diverse cooling effects in the early phase.
Fig. \ref{fig:kT2} shows the $kT_{\pm}$ evolution  
but for the ``High-$P_{0}$'' case. The initial temperature is given by
\begin{eqnarray}\label{eq:kT02}
kT_{\pm,0}
&=& 13~{\rm GeV}
\left(\frac{{\dot M}_{\rm j}}{10^{-3}~{\rm M_{\odot}yr^{-1}}}\right)^{-1}
\left(\frac{P_{0}V_{0}/t_{0}}{1.4\times 10^{48} {\rm erg~s^{-1}}}\right) . \nonumber \\
\end{eqnarray}
The behavior is essentially the same as  Fig. \ref{fig:kT}.
Adiabatic loss phases for the ``High-$P_{0}$'' case  
last longer than those for the ``Low-$P_{0}$'' case  
simply because of its higher $kT_{\pm,0}$.

Next, we show how $kT_{\pm,0}$
will change when we choose
different initial conditions.
The terms  ${\dot M}_{\rm j}$ and
$P_{0}V_{0}/t_{0}$ 
are the only ingredients which determine the initial
temperature $kT_{\pm,0}$.
The predicted $kT_{\pm,0}$ will increase
by the increase of 
$R_{0}$, $v_{0}$, and
$\rho_{\rm ext}$
according to $P_{0}V_{0}/t_{0}
\propto \rho_{\rm ext}R_{0}^{2}v_{0}^{3}$, and vice versa.
Therefore, in principle, 
$kT_{\pm,0}$ can become higher or lower than 
MeV/GeV ranges examined in Fig. \ref{fig:kT}.
When $kT_{\pm,0}$ is order of keV, 
the predicted $kT_{\pm,0}$ is smaller than 
$\Gamma_{\rm j}m_{e}c^{2}/3=1(\Gamma_{\rm j}/10)~{\rm MeV}$.
Therefore, the lobe plasma is gradually heated by
the jet injection and  $kT_{\pm}$ asymptotically
approaches to 
$\Gamma_{\rm j}m_{e}c^{2}/3=1(\Gamma_{\rm j}/10)~{\rm MeV}$. 
However, 
the case of ${\dot M}_{\rm j}\sim 10~{\rm M_{\odot}~yr^{-1}}$
requires $L_{\rm j}\sim 10^{48}~{\rm erg~s^{-1}}$ (see Eq. (\ref{eq:lkin}))
which tend to exceed the ones for very powerful AGN jets 
(e.g., Rawlings and Saunders 1991; Ito et al. 2008). 
Here we do not discuss this case further.

Lastly we comment on the formation of initial lobes. 
In order to accomplish the initial hot lobes at $t=t_{0}$, 
a heating process for $e^{\pm}$ pairs is needed at $t\le t_{0}$.
 Theoretically, a mechanism of electron heating (and acceleration)
 in a collisionless shock is a matter of debate 
 (e.g., Shimada and Hoshino 2000; Ohira and Takahara 2007).
 If pairs are  effectively heated up by various kind of plasma
 instabilities via hotter protons (e.g., Lyubarsky 2006),
 it could be possible to attain $kT_{\pm,0}\sim kT_{p,0}$.
 It is beyond the scope of this paper to explore microscopic 
 processes in the plasma. Instead, we  have uniquely determined $T_{\pm,0}$ 
 by using the quantities required from the observations and the 
 assumption of thermal pair expansions.
We have neglected the partial pressure of 
proton in $P_{0}$ merely for simplicity.
However the neglect does not affect the main arguments in this work. 
Because the protons also simply cool down 
by the adiabatic loss for $t>t_{0}$. 
To form ``High-$P_{0}$" lobe, 
very fast $\Gamma_{\rm j}\sim 50-100$ is required
when assuming that hot protons with $kT_{p}=\Gamma_{\rm j}m_{p}c^{2}/3$ 
are the heating source of $e^{\pm}$ pairs.
The required $\Gamma_{\rm j}$ is by a factor of $\sim 2-3$ larger than
the fastest $\Gamma_{\rm j}\sim 30$ estimated from 
VLBI observations (Kellermann et al. 2004)
and it is comparable to $\Gamma_{\rm j}\sim 50-100$ 
indicated by rapid variabilities of blazars
(e.g., Begelman et al. 2008; Ghisellini and Tavecchio 2008).
Since we take the oversimplified assumption of proton neglect,
it is not possible any further to explore the hot lobe formation at $t<t_{0}$.
We keep this as a subject for future investigation.
It is worth to mention that 
youngest radio sources
termed ``high frequency peakers" (HFPs) with ages
of $ < 10^{3}~{\rm yr}$
have been recently observed (e.g., Orienti et al. 2007)
and HFPs may give us substantial hints 
for the physics in $t \le t_{0}$.

\section{$\gamma$-ray emission from radio lobes}

\begin{figure}
\includegraphics[width=8cm]{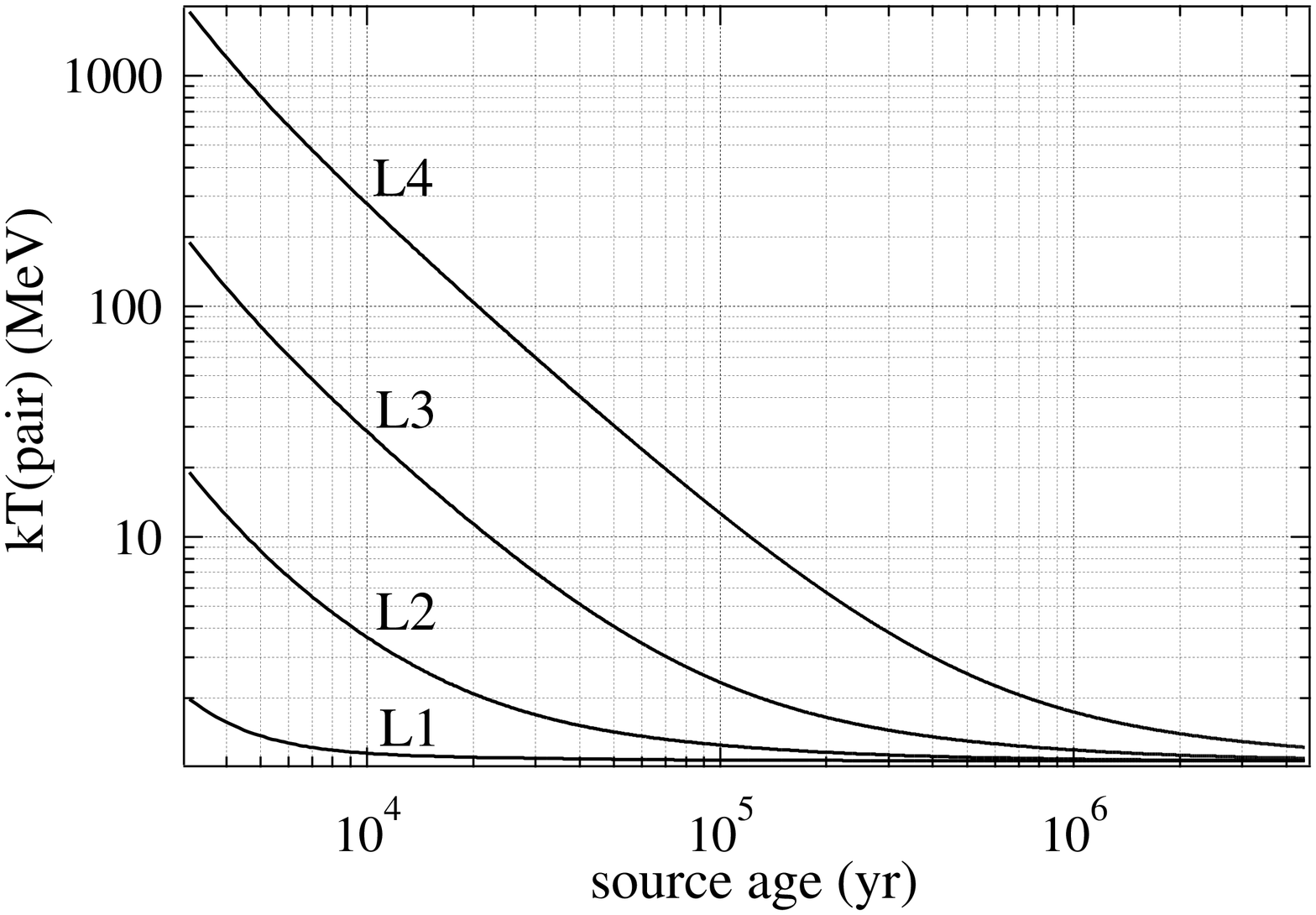}
\caption
{Evolution of $kT_{\pm}$ 
for the ``Low-$P_{0}$'' case 
which starts from the age of $t_{0}=3000~{\rm yr}$.
The examined  cases are 
${\dot M}_{\rm j}/(M_{\odot}~ {\rm yr^{-1}})
=10^{-4}$, $10^{-3}$, $10^{-2}$, and $10^{-1}$
and they are labeled as L4, L3, L2, and L1, respectively.
Corresponding kinetic power of the jet are, respectively, 
$L_{\rm j}=5.7\times 10^{42}~{\rm erg~s^{-1}} $,
$L_{\rm j}=5.7\times 10^{43}~{\rm erg~s^{-1}} $,
$L_{\rm j}=5.7\times 10^{44}~{\rm erg~s^{-1}} $, and 
$L_{\rm j}=5.7\times 10^{45}~{\rm erg~s^{-1}} $.
The temperature decrease by the adiabatic loss
are seen in the initial phase. 
Each case asymptotically goes to   
constant $kT_{\pm}$ phase which is predicted by KKI07.
Bremsstrahlung cooling is found to be negligible.}
\label{fig:kT}
\end{figure}

\begin{figure}
\includegraphics[width=8cm]{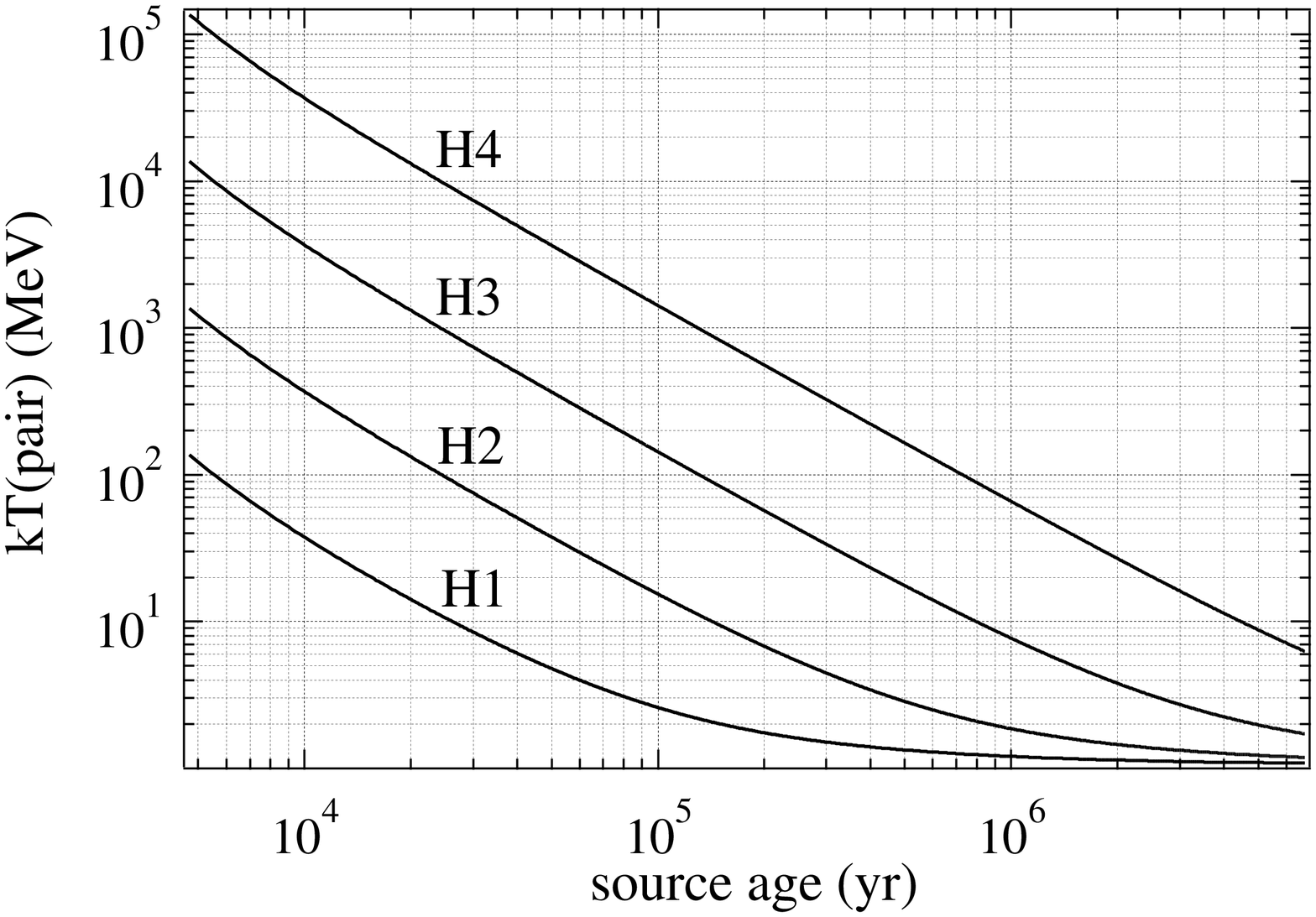}
\caption
{Evolution of $kT_{\pm}$ for the ``High-$P_{0}$'' case
which starts from the age of $t_{0}=4500~{\rm yr}$.
We examine the cases of 
${\dot M}_{\rm j}/(M_{\odot}~ {\rm yr^{-1}})
=10^{-4}$, $10^{-3}$, $10^{-2}$, and $10^{-1}$ and
they are labeled as H4, H3, H2, and H1, respectively.
Correspondingly, kinetic power of injected jet into
the lobes are, respectively, 
$L_{\rm j}=5.7\times 10^{42}~{\rm erg~s^{-1}} $,
$L_{\rm j}=5.7\times 10^{43}~{\rm erg~s^{-1}} $,
$L_{\rm j}=5.7\times 10^{44}~{\rm erg~s^{-1}} $, and 
$L_{\rm j}=5.7\times 10^{45}~{\rm erg~s^{-1}} $.
Similar to the ``Low-$P_{0}$'' case, 
adiabatic loss is dominant in the initial phase and
bremsstrahlung cooling is negligible.}
\label{fig:kT2}
\end{figure}

\begin{figure}
\includegraphics[width=8cm]{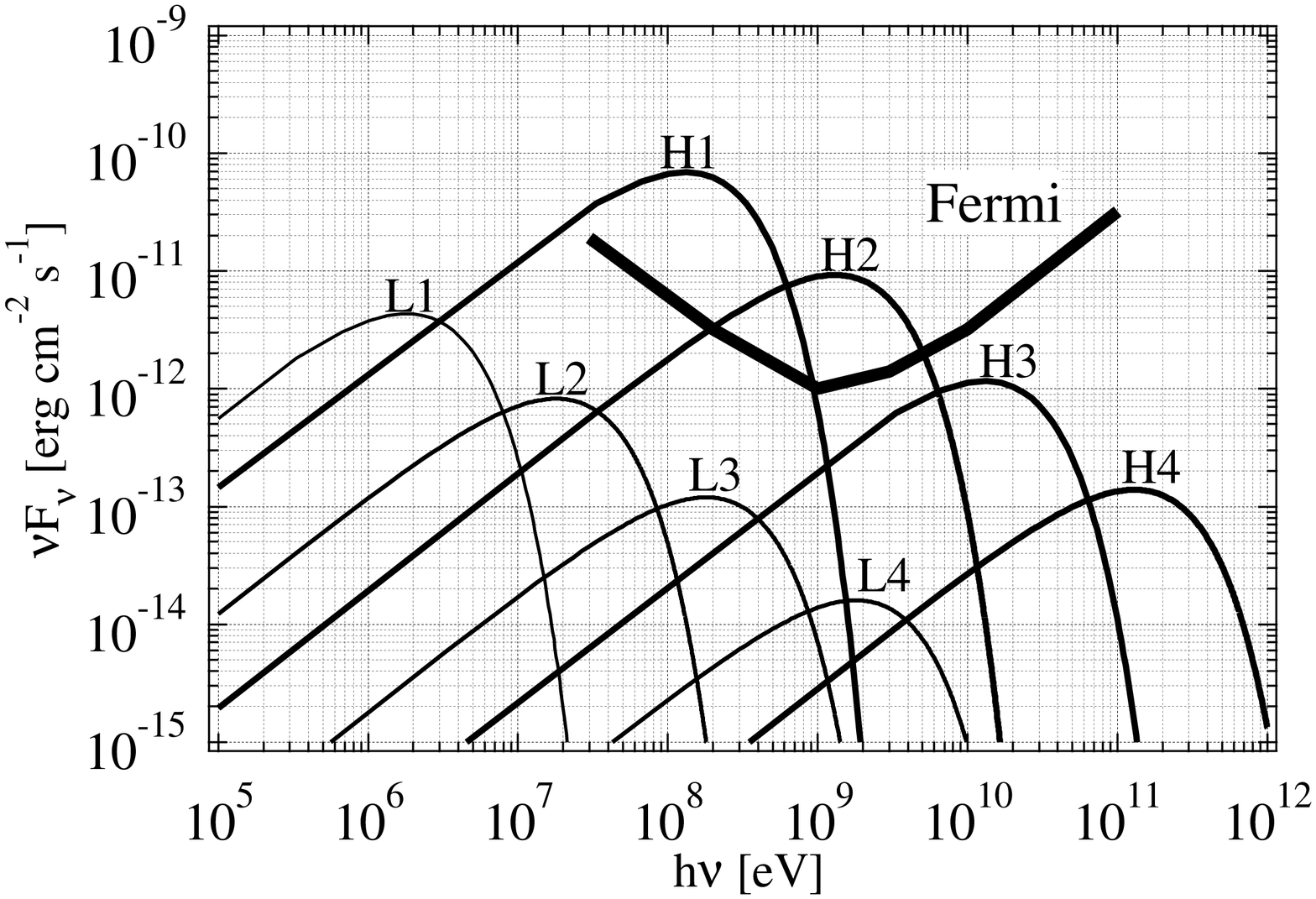}
\caption
{Predicted bremsstrahlung
emission from young radio lobes at $t=t_{0}$
with the distance $10^{2}$ Mpc. 
The spectra for ``High-$P_{0}$ '' case (H1, H2, H3, and H4) are
brighter than the sensitivity of Fermi Gamma-ray Space Telescope
while the spectra for ``Low-$P_{0}$'' case (L1, L2, L3, and L4) 
are less luminous for the detection.}
\label{fig:GeVlobe}
\end{figure}

Using Eqs. (\ref{eq:initial2}), (\ref{eq:kT0}), and (\ref{eq:kT02}),
we can estimate the bolometric bremsstrahlung luminosity  
$L_{\rm brem,0}=\epsilon_{\rm brem,0}V_{0}$ at $t=t_{0}$ for
the ``Low $P_{0}$'' case
\begin{eqnarray}\label{eq:l_brem}
L_{\rm brem,0} &=& 1.5 \times 10^{41}   
\left(\frac{n_{-,0}}{2.7 \times 10^{-2}~{\rm cm^{-3}}}\right)^{2}
\left(\frac{V_{0}}{1.2\times 10^{62}~{\rm cm^{3}}}\right) \nonumber \\
&& 
\left( \frac{kT_{\pm,0}}{0.18~{\rm GeV}}\right)
\left[1+
\ln\left(0.17 \frac{kT_{\pm,0}}{0.18~{\rm GeV}}\right)\right]~{\rm erg \ s^{-1}}~  
\end{eqnarray}
for ${\dot M}=10^{-3}~{\rm M_{\odot}~yr^{-1}}$.
The lobe for the ``High $P_{0}$'' case 
with the same ${\dot M}=10^{-3}~{\rm M_{\odot}~yr^{-1}}$
have hotter $e^{\pm}$ pairs 
and smaller  $n_{-,0}$ in larger $V_{0}$.
Then, the ``High $P_{0}$''case 
leads to brighter $L_{\rm brem,0}$ of
\begin{eqnarray}\label{eq:l_brem2}
L_{\rm brem,0} &=&  1.4 \times 10^{42}  
\left(\frac{n_{-,0}}{1.5 \times 10^{-3}~{\rm cm^{-3}}}\right)^{2}
\left(\frac{V_{0}}{3.3\times 10^{63}~{\rm cm^{3}}}\right) \nonumber \\
&& 
\left( \frac{kT_{\pm,0}}{13~{\rm GeV}}\right)
\left[1+
\ln\left(0.10\frac{kT_{\pm,0}}{13~{\rm GeV}}\right)\right]
~{\rm erg \ s^{-1}} . 
\end{eqnarray}
Fig. \ref{fig:GeVlobe} shows the  
bremsstrahlung spectra at $t=t_{0}$. 
Adopted  ${\dot M}_{\rm j}$ are same as in Fig. 1.
The source is located at the distance of
$10^{2}~{\rm Mpc}$ which corresponds to nearby 
observed CSO samples (e.g., Snellen et al. 2004).
For these cases, the predicted spectra are 
brighter than the sensitivity of 
Fermi/LAT (http://www-glast.stanford.edu/).
In other words,
we predict radio lobes in CSOs
as a new population of GeV-$\gamma$ emitter in the universe.
From Eqs. (\ref{eq:l_brem}) and  (\ref{eq:l_brem2}), we find that
the increase of ${\dot M}_{\rm j}$ leads to 
the enhancement of $L_{\rm brem,0}$ since 
$L_{\rm brem,0}\propto {\dot M}_{\rm j}$ holds
where we neglect the logalithm terms 
in (\ref{eq:l_brem}) and  (\ref{eq:l_brem2}). 
We can also say that lower $kT_{\pm,0}$ leads 
to brighter $L_{\rm brem,0}$ because 
$n_{\pm,0}^{2}\propto {\dot M}_{\rm j}^{2}$ dominates  
$kT_{\pm,0}\propto {\dot M}_{\rm j}^{-1}$, and vice versa.
Regarding the luminosity evolution, 
we obtain
$L_{\rm brem}\propto kT_{\pm}n_{\pm}^{2}V
\propto n_{\pm}$
where we use the relation of 
$kT_{\pm}\propto t^{-1}$ and $nV\propto t$ for the early phase.
We thus find that
$L_{\rm brem}$ decrease with time and it will fade out
because the lobe becomes dilute.
This sort of negative luminosity evolution has been well known
for synchrotron emission (Readhead et al. 1996b; Begelman 1996)
and bremsstrahlung emission (KKI07). 
Once we obtain $L_{\rm brem,0}$,
we can easily check that the radiative cooling is 
effective or not.
Since
$L_{\rm brem}$ decreases with time,
the effect of bremsstrahlung is most significant 
at $t=t_{0}$.
At the time, $L_{\rm brem,0}\ll L_{\rm j}$ holds 
where  $L_{\rm j}$ is the source term of the internal 
energy flux injected in the lobe.
Therefore we find that back-reaction
of bremsstrahlung cooling is negligible
and the behavior of $kT_{\pm}$ in Figs. \ref{fig:kT} and
\ref{fig:kT2} are governed by the adiabatic loss.

Lastly we discuss the change of
$L_{\rm brem,0}$ for different choices of
initial conditions. 
Two ingredients to determine $L_{\rm brem,0}$
are  $n_{-,0}$ and $kT_{\pm,0}$
for given $V_{0}$.
Concerning
$kT_{\pm,0}\propto P_{0}V_{0}/t_{0}$, 
the predicted $L_{\rm brem}$ increases
as $R_{0}$, $v_{0}$, and $\rho_{\rm ext}$ becomes larger
according to $kT_{\pm,0}\propto \rho_{\rm ext}R_{0}^{2}v_{0}^{3}$,
and vice versa.
Hence 
it is worth to note that lobes with faster 
sideways expansions will shine  brighter. 
Indeed there are some CSOs with $v_{0}>0.1c$ 
(Polatidis and Conway 2003; 
Gugliucci et al. 2005, 2007).
The quantity $n_{-,0}$ is the  
important parameter, since 
$n_{-,0}$ dependence of $L_{\rm brem,0}$
is significant as $L_{\rm brem,0}\propto n_{-,0}^{2}$.
For the estimate of $n_{-,0}$, 
we assumed the $e^{\pm}$ pair jet so far. 
If instead we assume the pure electron/proton jet
for given $L_{\rm j}$ and $\Gamma_{\rm j}$, 
then $n_{-,0}$ decreases by a factor of $m_{e}/m_{p}$.
Then $L_{\rm brem,0}$ decreases by a
factor of $(m_{e}/m_{p})^{2}$,
and the predicted $L_{\rm brem,0}$ becomes
much smaller than the Fermi/LAT sensitivity.


\section{Summary and Discussion}

We have investigated the temperature and luminosity evolutions 
of radio lobes of CSOs which expands by their own thermal pressure.
In the previous work of KKI07, 
MeV $\gamma$-ray bremsstrahlung emission was predicted from
young lobes in the regime of no coolings.
In this work, 
we include cooling effects of bremsstrahlung emission
and adiabatic loss.
Below we summarize  the main results of the present work.

\begin{enumerate}

\item

We examine the evolution of $kT_{\pm}$
together with 
 the initial conditions determined by observed CSOs. 
By solving a set of equations describing 
the expanding lobe evolution,
it is found that the lobes initially have electron temperature of $\sim$GeV 
for $\Gamma_{\rm j}\sim 10$,
the lobes then cool down to MeV by the adiabatic loss.
During the early phase,  $kT_{\pm}$ is
governed by the adiabatic loss alone. Since 
the adiabatic loss is more effective than bremsstrahlung cooling
in any  case.
In the late phase, $kT_{\pm}$ asymptotically
approaches to  the constant temperature of
$\Gamma_{\rm j}m_{e}c^{2}/3=
1~(\Gamma_{\rm j}/10)~{\rm MeV}$ which has been predicted in KKI07.

\item

Thermal bremsstrahlung 
emission peaked about GeV-$\gamma$ band is 
newly predicted in CSOs 
because $n_{\pm}$ and $kT_{\pm}$ in younger radio lobes 
are larger than those in older ones. 
These spectra can be
brighter than the sensitivity of Fermi/LAT for nearby CSOs
in the case of High-$P_{0}$ lobe.
This means that young radio-lobes 
can be a new population of GeV-$\gamma$ emitter in the universe.
From Eq. (\ref{eq:l_brem}) showing $L_{\rm brem,0}$,
we see that the radio lobes with larger $v_{0}$ 
(i.e., faster expansions)
and/or the one with larger $n_{-}$ (i.e., larger $L_{\rm j}$)
yield brighter emission for given  $R_{0}$.
As for a specific source, we stress the importance of
young recurrent lobe 3C 84 (e.g., Asada et al. 2006). 
Since 3C 84 is located at the center of 
very nearby Perceus cluster ($z=0.018$), a deep observation of the 
Fermi/LAT collabolated with radio obserbations on it 
will give us tight constraints on the physics of young radio lobes.
Near future mission 
of the VLBI Space Observatory Programme 2
(VSOP-2) with unprecedented high angular resolution 
(Tsuboi et al. 2009)
would play important roles for 
direct constraint on $v_{0}$ and $R_{0}$ of HFPs.

\item

Stawarz et al. (2008) investigated 
a variety of non-thermal $\gamma$-ray 
emissions from lobes of CSOs.
They claim that the predicted non-thermal emissions 
can be also detected by Fermi/LAT with an assumed certain 
electron acceleration efficiency.
In GeV $\gamma$ range, the dominant components
are inverse-Compton scattered emissions
of ultraviolet  and infrared photons.
It is worth to note that 
even though the luminosity of thermal bremsstrahlung
and inverse-Compton ones are comparable,
the spectrum shape thermal component 
is quite different from non-thermal spectra. 
Hence it is straightforward to distinguish whether 
the emission is thermal or non-thermal one.

\item

The observations of Fermi/LAT will be tests for
some unresolved questions of AGN jets.
Suppose the case that we exactly know  
$R_{0}$,
$v_{0}$, and 
$n_{\rm ext}$, and
a source distance.
If we observe the predicted thermal 
GeV-$\gamma$ emission from CSOs,
then it suppose the scenario in which CSOs have
relativistic jets and their lobes thermally expand.
If we do not detect it, it is attribute to
lower $kT_{\pm}$ and/or smaller  $n_{\pm}$.
Possible reasons are as follows;
(1) the jet is mainly made of $e/p$ plasma with the same $L_{\rm j}$,
(2) the jet consists of  $e^{\pm}$ plasma on the whole
but with smaller $L_{\rm j}$,
(3) the lobe is expanded by energetic
non-thermal particle and the actual $n_{\pm}$ is  smaller,
(4) the jet has non-relativistic speed 
    which leads to the lower $kT_{\pm}$, 
(5) other cooling processes could make $kT_{\pm}$  lower.
We remain them as our future investigations.

\end{enumerate}

\section*{Acknowledgments}
We thank J. Kataoka, the referee, whose beneficial comments
helped us to substantially improve the paper.
We are indebted to C. R. Kaiser and M. Sikora
for valuable comments.
NK is supported by Grant-in-Aid for JSPS Fellows.
HI acknowledge
the Grant for Special Research Projects at Waseda University.

\end{document}